\documentclass[a4,11pt]{amsart}
\usepackage{amssymb}
\usepackage{amsmath}
\usepackage{amsfonts}
\usepackage{amsmath,amssymb,pstricks,pst-plot,pst-node,psfrag,graphicx,pst-grad,amsfonts}
\usepackage{float}
\usepackage{enumerate}

\begin{document}

\title{Transmuted Generalized Inverse Weibull Distribution}

\author[Faton Merovci]{Faton Merovci  }
\address{Faton Merovci
\newline \indent Department of Mathematics,
\newline \indent University of Prishtina "Hasan Prishtina",
\newline \indent Republic of Kosovo}

\email{fmerovci@yahoo.com}

\author[Ibrahim Elbatal]{Ibrahim Elbatal  }
\address{Ibrahim Elbatal
\newline \indent Institute of Statistical Studies and Research,
\newline \indent Department of Mathematical Statistics,
\newline \indent Cairo University}

\email{i\_elbatal@staff.cu.edu.eg}

\author[Alaa  Ahmed ]{Alaa  Ahmed }
\address{Alaa  Ahmed 
\newline \indent Institute of Statistical Studies and Research,
\newline \indent Department of Applied Statistics and Econometrics
\newline \indent Cairo University
}
\email{Alaa\_mnn@yahoo.com}
\maketitle
\begin{abstract}
A generalization of the generalized inverse Weibull distribution so-called
transmuted generalized inverse Weibull distribution is proposed and studied.
We will use the quadratic rank transmutation map (QRTM) in order to generate
a flexible family of probability distributions taking generalized inverse
Weibull distribution as the base value distribution by introducing a new
parameter that would offer more distributional flexibility. Various
structural properties including explicit expressions for the moments,
quantiles, and moment generating function of the new distribution are
derived.We proposed the method of maximum likelihood for estimating the
model parameters and obtain the observed information matrix.
A real data set  are used to compare the ﬂexibility of the transmuted  version
versus the generalized inverse Weibull distribution.

\end{abstract}

\section{Introduction}

The inverse Weibull distribution is another life time probability
distribution which can be used in the reliability engineering discipline.
The inverse Weibull distribution can be used to model a variety of failure
characteristics such as infant mortality, useful life and wear- out periods.
It can also be used to determine the cost effectiveness , maintenance
periods of reliability centered maintenance activities and applications in
medicine, reliability and ecology. Keller (1985) obtained the inverse
Weibull model by investigating failures of mechanical components subject to
degradation. Drapella (1993), Mudholkar and Kollia (1994) and Gusm\~{a}o et
al. (2011) introduced the generalized inverse Weibull distribution, among
others. The cumulative distribution function (cdf) of the generalized
inverse Weibull (GIW) distribution can be defined by
\begin{equation}\label{eq1.1}
G(x,\alpha ,\gamma ,\theta )=e^{-\gamma (\alpha x)^{-\beta }},\alpha
>0,\gamma >0,\beta >0\text{ and }x\geq 0.
\end{equation}%
where\ $\alpha $ is scale parameter and $\beta , \gamma $ are shape
parameters respectively. The corresponding probability density function
(pdf) is given by%
\begin{equation}\label{eq1.2}
g(x,\alpha ,\gamma ,\theta )=\alpha \beta \gamma (\alpha x)^{-\beta
-1}e^{-\gamma (\alpha x)^{-\beta }}.
\end{equation}%
In this article we present a new generalization of generalized inverse
Weibull distribution called the transmuted generalized inverse Weibull
distribution. We will derive the subject distribution using the quadratic
rank transmutation map studied by Shaw et al. (2007). A random variable $X$
is said to have transmuted distribution if its cumulative distribution
function(cdf) is given by
\begin{equation}\label{eq1.3}
F(x)=(1+\lambda )G(x)-\lambda G(x)^{2},\left\vert \lambda \right\vert \leq 1,
\end{equation}%
where $G(x)$ is the cdf of the base distribution, which on differentiation
yields,%
\begin{equation}\label{eq1.4}
f(x)=g(x)\left[ (1+\lambda )-2\lambda G(x)\right]
\end{equation}%
where $f(x)$ and $g(x)$ are the corresponding pdfs associated with cdf $F(x)$
and $G(x)$ respectively. An extensive information about the quadratic rank
transmutation map is given in Shaw et al. (2007). Observe that at $\lambda
=0 $ we have the distribution of the base random variable.

Many authors dealing with the generalization of some well- known
distributions. Aryal and Tsokos (2009) defined the transmuted generalized
extreme value distribution and they studied some basic mathematical
characteristics of transmuted Gumbel probability distribution and it has
been observed that the transmuted Gumbel can be used to model climate data.
Also Aryal and Tsokos (2011) presented a new generalization of Weibull
distribution called the transmuted Weibull distribution . Recently, Aryal
(2013) proposed and studied the various structural properties of the
transmuted Log- Logistic distribution. and Khan and King (2013)
introduced the transmuted modified Weibull distribution which extended
recent development on transmuted Weibull distribution by Aryal et al.
(2011), and they studied the mathematical properties and maximum likelihood
estimation of the unknown parameters. In the present study we will provide
mathematical formulation of the transmuted generalized inverted exponential
distribution and some of its properties. We will also provide possible area
of applications.

The rest of the paper is organized as follows. In Section 3 we demonstrate
transmuted probability distribution.  In Section 4 , we find the reliability
functions of the subject model. The statistical properties include quantile
functions, moments and moment generating functions are derived in Section 5.The minimum, maximum and median order statistics models are discussed in
Section 6. Least Squares and Weighted Least Squares Estimators are discused in section 7.Section 8 we demonstrate the maximum likelihood
estimates and the asymptotic confidence intervals of the unknown parameters. In
section 9, the TGIW distribution is applied to a real data set. Finally, in Section 10, we provide some conclusion.

\section{Transmutation Map}

In this section we demonstrate transmuted probability distribution. Let $F_{1}$ and $F_{2}$ be the cumulative distribution functions, of two
distributions with a common sample space. The general rank transmutation as
given in (2007) is defined as%
\begin{equation*}
G_{R12}(u)=F_{2}(F_{1}^{-1}(u))\text{ \ and }G_{R21}(u)=F_{1}(F_{2}^{-1}(u)).
\end{equation*}%
Note that the inverse cumulative distribution function also known as
quantile function is defined as
\begin{equation*}
F^{-1}(y)=\text{ inf}_{x\in R}\left\{ F(x)\geq y\right\} \text{ for }y\in %
\left[ 0,1\right] .
\end{equation*}%
The functions $G_{R12}(u)$ and $G_{R21}(u)$ both map the unit interval $I=$ $%
\left[ 0,1\right] $ into itself, and under suitable assumptions are mutual
inverses and they satisfy $G_{Rij}(0)=0$ and $G_{Rij}(0)=1.$ A quadratic Rank
Transmutation Map (QRTM) is defined as%
\begin{equation}\label{eq2.1}
G_{R12}(u)=u+\lambda u(1-u),\left\vert \lambda \right\vert \leq 1,
\end{equation}%
from which it follows that the cdf's satisfy the relationship%
\begin{equation}\label{eq2.2}
F_{2}(x)=(1+\lambda )F_{1}(x)-\lambda F_{1}(x)^{2}
\end{equation}%
which on differentiation yields,%
\begin{equation}\label{eq2.3}
f_{2}(x)=f_{1}(x)\left[ (1+\lambda )-2\lambda F_{1}(x)\right]
\end{equation}%
where $f_{1}(x)$ and $f_{2}(x)$ are the corresponding pdfs associated with
cdf $F_{1}(x)$ and $F_{2}(x)$ respectively. An extensive information about
the quadratic rank transmutation map is given in Shaw et al. (2007). Observe
that at $\lambda =0$ we have the distribution of the base random variable.
The function $f_{2}(x)$ in given \eqref{eq2.3}
satisfies the property of probability density function.

\section{Transmuted Generalized Inverse Weibull Distribution}

In this section we studied the transmuted generalized inverse weibull (TGIW)
Distribution and the sub-models of this distribution. Now using \eqref{eq1.1} and
\eqref{eq1.2} we have the cdf of transmuted generalized inverted exponential
distribution
\begin{equation}\label{eq2.1}
F_{TGIW}(x)=e^{-\gamma (\alpha x)^{-\beta }}\left[ 1+\lambda -\lambda
e^{-\gamma (\alpha x)^{-\beta }}\right]
\end{equation}
where\ $\alpha $ is scale parameter and $\beta , \gamma $ are shape
parameters representing the different patterns of the transmuted generalized
inverse weibull distribution\ and $\lambda $ is the transmuted parameter.The
corresponding probability density function (pdf) of the transmuted
generalized inverse Weibull distribution is given by%
\begin{equation}\label{eq2.2}
f_{TGIW}(x)=\alpha \beta \gamma (\alpha x)^{-\beta -1}e^{-\gamma (\alpha
x)^{-\beta }}\left[ 1+\lambda -2\lambda e^{-\gamma (\alpha x)^{-\beta }}
\right] .
\end{equation}

Figure 1 and 2 illustrates some of the possible shapes of the pdf  and cdf of TGIW distribution  for selected values of the parameters $ \beta, \gamma $ and $\lambda,$ by keeping $alpha=1$, respectively.

\begin{figure}[H]
\centering
  \includegraphics[width=8.0cm]{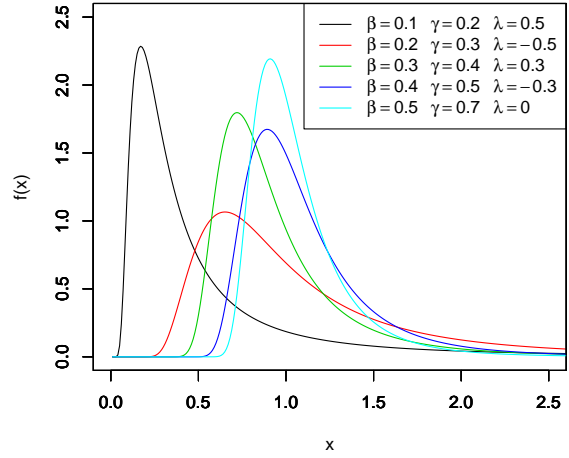}
 \caption{The pdf's of various TGIW distributions.\label{fig1.pdf}}
 \end{figure}

\begin{figure}[H]
\centering
  \includegraphics[width=8.0cm]{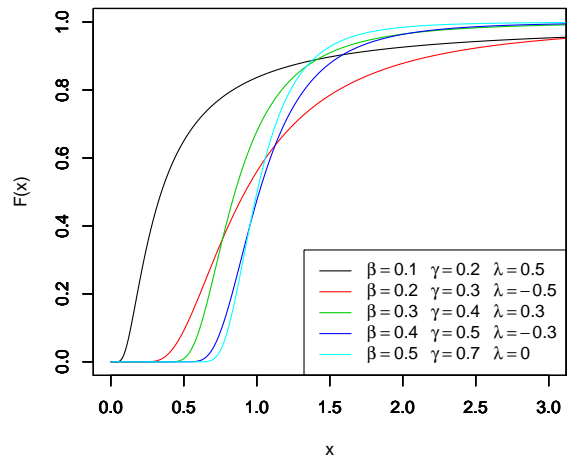}
 \caption{The cdf of various TGIW distributions.\label{fig2.pdf}}
 \end{figure}

The transmuted generalized inverse Weibull distribution is very flexible
model that approaches to different distributions when its parameters are
changed. The flexibility of the transmuted generalized inverse Weibull
distribution is explained in the following . If $X$ is a random variable
with pdf \eqref{eq2.2}, then we have the following cases.

\begin{enumerate}[(a)]

\item If $\gamma =1,$we get the transmuted inverse Weibull.

\item If $\lambda =0$ and $\gamma =1,$we get the inverse Weibull.

\item If $\beta =1$ , $\gamma =1$we get the transmuted inverse exponential
distribution.

\item If $\beta =1$ , $\gamma =1$ and $\lambda =0$ we get the inverse
exponential distribution.

\item If $\beta =2$ , $\gamma =1$ we get transmuted inverse Rayleigh
distribution.

\item If $\beta =2$ , $\gamma =1$ and $\lambda =0$ we get the inverse Rayleigh
distribution.

\item If $\alpha =1$ we get the transmuted Frechet distribution.

\item If $\alpha =1$ and and $\lambda =0$ we get Frechet distribution.
\end{enumerate}
\section{Reliability Analysis}

The reliability function $R(x)$, which is the probability of an item not
failing prior to some time $t$, is defined by $R(x)=1-F(x)$. The reliability
function of a transmuted generalized inverse Weibull distribution $%
R_{TGIW}(x)$, it can be a useful characterization of life time data
analysis.
\begin{align}\label{eq3.1}
R_{TGIW}(x) &=1-F_{TGIW}(x)  \notag \\
&=1-e^{-\gamma (\alpha x)^{-\beta }}\left[ 1+\lambda -\lambda e^{-\gamma
(\alpha x)^{-\beta }}\right] .
\end{align}%
It is important to note that $R_{TGIW}(x)+F_{TGIW}(x)=1$ . The
other characteristic of interest of a random variable is the hazard rate
function defined by $h_{TGIW}(x)=\frac{f_{TGIW}(x)}{1-F_{TGIW}(x)}$ which is
an important quantity characterizing life phenomenon. It can be loosely
interpreted as the conditional probability of failure, given it has survived
to the time $t$. The hazard rate function for a transmuted generalized
inverse Weibull distribution is defined by
\begin{align}\label{eq3.2}
h_{TGIW}(x) &=\frac{f_{TGIW}(x)}{1-F_{TGIW}(x)}  \notag \\
&=\frac{\alpha \beta \gamma (\alpha x)^{-\beta -1}e^{-\gamma (\alpha
x)^{-\beta }}\left[ 1+\lambda -2\lambda e^{-\gamma (\alpha x)^{-\beta }}%
\right] }{1-e^{-\gamma (\alpha x)^{-\beta }}\left[ 1+\lambda -\lambda
e^{-\gamma (\alpha x)^{-\beta }}\right] .}
\end{align}%

Figure 3 and 4 illustrates some of the possible shapes of the hazard rate function  and survival function of TGIW distribution  for selected values of the parameters $ \beta, \gamma $ and $\lambda,$ by keeping $alpha=1$, respectively.

\begin{figure}[H]
\centering
  \includegraphics[width=8.0cm]{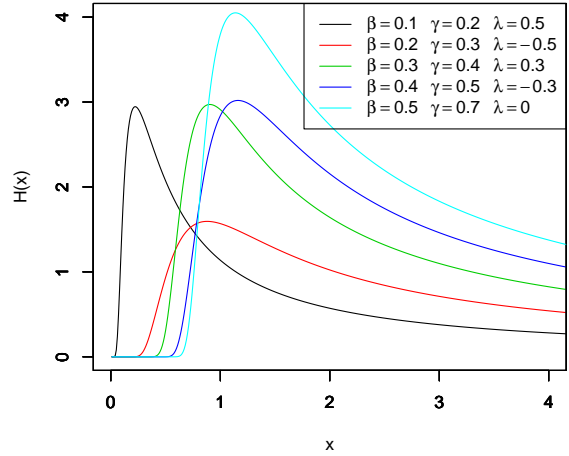}
 \caption{The hazard function of various TGIW distributions\label{fig3.pdf}}
 \end{figure}

\begin{figure}[H]
\centering
  \includegraphics[width=8.0cm]{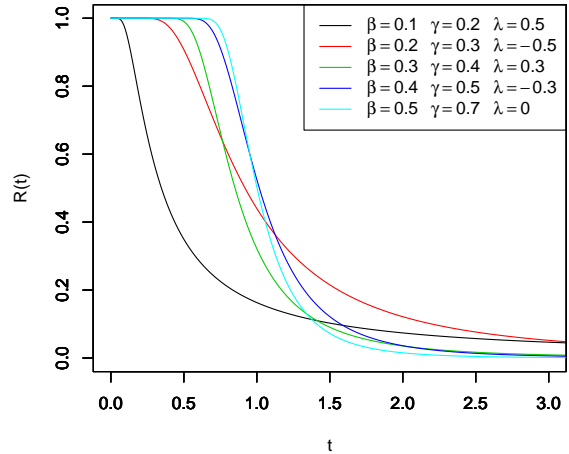}
 \caption{The survival function of various TGIW distributions\label{fig4.pdf}}
 \end{figure}
It is important to note that the units for $h_{TGIW}(x)$ is the probability
of failure per unit of time, distance or cycles. These failure rates are
defined with different choices of parameters.The cumulative hazard function
of the transmuted generalized inverse Weibull distribution is denoted by
\begin{equation}\label{eq3.3}
H_{TGIW}(x)=-\ln \left\vert e^{-\gamma (\alpha x)^{-\beta }}\left[ 1+\lambda
-\lambda e^{-\gamma (\alpha x)^{-\beta }}\right] \right\vert
\end{equation}
It is important to note that the units for $H_{TGIW}(x)$ is the cumulative
probability of failure per unit of time, distance or cycles.

\section{Statistical Properties}

In this section we discuss the statistical properties of the transmuted
generalized inverse Weibull distribution. Specifically quantile and random
number generation function, moments and moment generating function.

\subsection{Quantile and Median}

The quantile $x_{q}$ of the $T_{GIW}$ $(\alpha ,\beta ,\gamma ,\lambda ,x)$
is real solution of the following equation%
\begin{equation}\label{eq4.1}
x_{q}=\left\{ \frac{-1}{\gamma }\ln \left( \frac{(\lambda +1)+\sqrt{(\lambda
+1)^{2}-4\lambda q}}{2\lambda }\right) \right\} ^{\frac{-1}{\beta }}
\end{equation}%
The above equation has no closed form solution in $x_{q}$, so we have to use
a numerical technique such as a Newton- Raphson method to get the quantile.
If we put $q=0.5$ in equation \eqref{eq4.1} one gets the median of $T_{GIW}(\alpha ,\beta ,\gamma ,\lambda ,x)$

\subsection{Random Number Generation}

The random number generation as $x$ of the $T_{GIW}(\alpha ,\beta ,\gamma,\lambda ,x)$ is defined by the following relation%
\begin{equation}\label{eq4.2}
x_{q}=\left\{ \frac{-1}{\gamma }\ln \left( \frac{(\lambda +1)+\sqrt{(\lambda
+1)^{2}-4\lambda \phi }}{2\lambda }\right) \right\} ^{\frac{-1}{\beta }}%
\text{\ where }\phi \sim U(0,1).
\end{equation}

\subsection{Moments}

The following theorem gives the $r_{th}$ moment $(\mu _{r}^{^{\prime }})$ of
the $T_{GIW}$ $(\alpha ,\beta ,\gamma ,\lambda ,x).$\medskip \newline

\textbf{Theorem (4.1)}.
\medskip
If $X$ has the $T_{GIW}(\alpha ,\beta ,\gamma ,\lambda ,x)$ with $\left\vert \lambda \right\vert \leq 1$, then the $r_{th}$ non central
moments are given by\medskip \newline
\begin{equation}\label{eq4.3}
\mu _{r}^{^{\prime }}(x)=E(X^{r})=\frac{\gamma ^{\frac{r}{\beta }}\Gamma (1-%
\frac{r}{\beta })}{\alpha ^{r}}\left[ 1+\lambda -\lambda (2)^{\frac{r}{\beta
}}\right] .
\end{equation}%
Proof:\medskip \newline
Starting with%
\begin{align}\label{eq4.4}
\mu _{r}^{^{\prime }}(x) &=\int\nolimits_{0}^{\infty }x^{r}f_{TGIW}(\alpha
,\beta ,\gamma ,\lambda ,x)dx  \notag \\
&=\int\limits_{0}^{\infty }x^{r}\alpha \beta \gamma (\alpha x)^{-\beta
-1}e^{-\gamma (\alpha x)^{-\beta }}\left[ 1+\lambda -2\lambda e^{-\gamma
(\alpha x)^{-\beta }}\right]   \notag \\
&=\left\{ \frac{\alpha \beta \gamma }{\alpha ^{r}}\left( 1+\lambda \right)
\int\nolimits_{0}^{\infty }(\alpha x)^{r-\beta -1}e^{-\gamma (\alpha
x)^{-\beta }}dx\right.  \notag \\
&-\left. \frac{2\lambda \alpha \beta \gamma }{\alpha ^{r}}%
\int\limits_{0}^{\infty }(\alpha x)^{r-\beta -1}e^{-2\gamma (\alpha
x)^{-\beta }}dx\right\}
\end{align}%
let $\gamma (\alpha x)^{-\beta }=t$  then $x=\frac{1}{\alpha }\gamma ^{
\frac{1}{\beta }}t^{\frac{-1}{\beta }},$ therefore
\begin{align}\label{eq4.5}
\mu _{r}^{^{\prime }}(x) &=\frac{\left( 1+\lambda \right) }{\alpha ^{r}}%
\gamma ^{\frac{r}{\beta }}\Gamma (1-\frac{r}{\beta })-\frac{\lambda }{\alpha
^{r}}(2\gamma )^{\frac{r}{\beta }}\Gamma (1-\frac{r}{\beta })  \notag \\
&=\frac{\gamma ^{\frac{r}{\beta }}\Gamma (1-\frac{r}{\beta })}{\alpha ^{r}}%
\left[ 1+\lambda -\lambda (2)^{\frac{r}{\beta }}\right] .
\end{align}%
Which completes the proof.

Based on Theorem (4.1) the coefficient of variation, coefficient of skewness
and coefficient of kurtosis of $T_{GIW}$ $(\alpha ,\beta ,\gamma ,\lambda
,x) $ distribution can be obtained according to the following relation%
\begin{align*}
CV_{TMIW} &=\sqrt{\frac{\mu _{2}}{\mu _{1}}-1}, \\
CS_{TMIW} &=\frac{\mu _{3}-3\mu _{2}\mu _{1}+2\mu _{1}^{3}}{(\mu _{2}-\mu
_{1})^{\frac{3}{2}}} \\
\text{and}CK_{TMIW} &=\frac{\mu _{4}-4\mu _{3}\mu _{1}+6\mu _{2}\mu _{1}^{2}%
}{(\mu _{2}-\mu _{1}^{2})^{2}}.
\end{align*}

\subsection{Moment Generating Function}

In this subsection we derived the moment generating function (mgf) of
transmuted generalized inverse Weibull distribution.\medskip \newline
\textbf{Theorem (4.2): }If $X$ has the$T_{GIW}$ $(\alpha ,\beta ,\gamma
,\lambda ,x)$ with $\left\vert \lambda \right\vert \leq 1$, then the the
moment generating function (mgf) of $X$ \ is given as follows%
\begin{equation}\label{eq4.6}
M_{X}(t)=\sum\limits_{r=0}^{\infty }\frac{t^{r}\gamma ^{\frac{r}{\beta }%
}\Gamma (1-\frac{r}{\beta })}{r!\alpha ^{r}}\left[ 1+\lambda -\lambda (2)^{%
\frac{r}{\beta }}\right] .
\end{equation}%
\textbf{Proof:}%
\begin{align}\label{eq4.7}
M_{X}(t) &=\int\nolimits_{0}^{\infty }e^{tx}f_{T_{GIW}}(\alpha ,\beta
,\gamma ,\lambda ,x)dx  \notag \\
&=\sum\nolimits_{r=0}^{\infty }\frac{t^{r}}{r!}\text{ }x^{r}f_{T_{GIW}}(%
\alpha ,\beta ,\gamma ,\lambda ,x)dx  \notag \\
&=\sum\nolimits_{r=0}^{\infty }\frac{t^{r}}{r!}\mu _{r}^{^{\prime }}(x)
\end{align}%
using equations \eqref{eq4.3} into relation \eqref{eq4.7} we get the following%
\begin{equation*}
M_{X}(t)=\sum\limits_{r=0}^{\infty }\frac{t^{r}\gamma ^{\frac{r}{\beta }%
}\Gamma (1-\frac{r}{\beta })}{r!\alpha ^{r}}\left[ 1+\lambda -\lambda (2)^{%
\frac{r}{\beta }}\right] .
\end{equation*}%
Which completes the proof.

\section{Order Statistics}

In fact, the order statistics have many applications in reliability and life
testing. The order statistics arise in the study of reliability of a system.
Let $X_{1},X_{2},...,X_{n}$ be a simple random sample from $T_{GIW}$ $%
(\alpha ,\beta ,\gamma ,\lambda ,x)$ with cumulative distribution function
and probability density function as in \eqref{eq2.1} and \eqref{eq2.2}, respectively. Let $X_{(1:n)}\leq X_{(2:n)}\leq \ldots\leq X_{(n:n)}$ denote the order
statistics obtained from this sample. In reliability literature, $X_{(i:n)}$
denote the lifetime of an $(n-i+1)-$ out$-$ of$-$ $n$ system which consists
of $n$ independent and identically components. Then the pdf of $X_{(i:n)}$ $%
,1\leq i\leq n$ is given by
\begin{equation}\label{eq5.1}
f_{i::n}(x)=\frac{1}{\beta (i,n-i+1)}\left[ F(x,\Phi )\right] ^{i-1}\left[
1-F(x,\Phi )\right] ^{n-i}f(x,\Phi )
\end{equation}%
where $\Phi =((\alpha ,\beta ,\gamma ,\lambda )$. Also, the joint pdf of $X_{(i:n)},X_{(j:n)}$ and $1\leq i\leq j\leq n$ is%
\begin{equation}\label{eq5.2}
f_{i:j:n}(x_{i},x_{j})=C\text{ }\left[ F(x_{i})\right] ^{i-1}\left[
F(x_{j})-F(x_{i})\right] ^{j-i-1}\left[ 1-F(x_{j})\right]
^{n-j}f(x_{i})f(x_{j})
\end{equation}%
where
\begin{equation*}
C=\frac{n!}{(i-1)!(j-i-1)!(n-j)!}
\end{equation*}%
We defined the first order statistics $X_{(1)}=Min(X_{1},X_{2},\ldots,X_{n}),$  the last order statistics as $X_{(n)}=Max(X_{1},X_{2},\ldots,X_{n}) $ and median order $X_{m+1}$ .

\subsection{Distribution of Minimum , Maximum and Median}

Let $X_{1},X_{2},\ldots,X_{n}$ be independently identically distributed
order random variables from the transmuted generalized inverse weibull
distribution having first , last and median order probability density
function are given by the following%
\begin{align}\label{eq5.3}
f_{1:n}(x) &=n\left[ 1-F(x,\Phi )\right] ^{n-1}f(x,\Phi )  \notag \\
&=n\left\{ 1-e^{-\gamma (\alpha x_{(1)})^{-\beta }}\left[ 1+\lambda
-\lambda e^{-\gamma (\alpha x_{(1)})^{-\beta }}\right] \right\} ^{n-1}
\notag \\
&\times \alpha \beta \gamma (\alpha x_{(1)})^{-\beta -1}e^{-\gamma (\alpha
x_{(1)})^{-\beta }}\left[ 1+\lambda -2\lambda e^{-\gamma (\alpha
x_{(1)})^{-\beta }}\right]
\end{align}%
\begin{align}\label{eq5.4}
f_{n:n}(x) &=n\left[ F(x_{(n)},\Phi )\right] ^{n-1}f(x_{(n)}),\Phi )  \notag \\
&=n\left\{ e^{-\gamma (\alpha x_{(n)})^{-\beta }}\left[ 1+\lambda -\lambda
e^{-\gamma (\alpha x_{(n)})^{-\beta }}\right] \right\} ^{n-1}  \notag \\
&\times \alpha \beta \gamma (\alpha x_{(n)})^{-\beta -1}e^{-\gamma (\alpha
x_{(n)})^{-\beta }}\left[ 1+\lambda -2\lambda e^{-\gamma (\alpha
x_{(n)})^{-\beta }}\right]
\end{align}
and
\begin{align}\label{eq5.5}
f_{m+1:n}(\widetilde{x}) &=\frac{(2m+1)!}{m!m!}(F(\widetilde{x}))^{m}(1-F(%
\widetilde{x}))^{m}f(\widetilde{x})  \notag \\
&=\frac{(2m+1)!}{m!m!}\left\{ e^{-\gamma (\alpha x_{(m+1)})^{-\beta }}\left[
1+\lambda -\lambda e^{-\gamma (\alpha x_{(m+1)})^{-\beta }}\right] \right\}
^{m}  \notag \\
&\times \left\{ 1-e^{-\gamma (\alpha x_{(m+1)})^{-\beta }}\left[ 1+\lambda
-\lambda e^{-\gamma (\alpha x_{(m+1)})^{-\beta }}\right] \right\} ^{m}
\notag \\
&\times \alpha \beta \gamma (\alpha x_{(m+1)})^{-\beta -1}e^{-\gamma
(\alpha x_{(m+1)})^{-\beta }}\left[ 1+\lambda -2\lambda e^{-\gamma (\alpha
x_{(m+1)})^{-\beta }}\right]
\end{align}

\subsection{Joint Distribution of the i$_{th}$ and j$_{th}$ order Statistics}

The joint distribution of the the $i_{th}$ and $j_{th}$ order statistics
from transmuted generalized inverse Weibull is%
\begin{align}\label{eq5.6}
f_{i:j:n}(x_{i},x_{j}) &=C\text{ }\left[ F(x_{i})\right] ^{i-1}\left[
F(x_{j})-F(x_{i})\right] ^{j-i-1}\left[ 1-F(x_{j})\right]
^{n-j}f(x_{i})f(x_{j})  \notag \\
&=C\left\{ h_{(i)}\left[ 1+\lambda -\lambda h_{(i)}\right] \right\} ^{i-1}
\notag \\
&\times \left\{ h_{(j)}\left[ 1+\lambda -\lambda h_{(j)}\right] -h_{(i)}%
\left[ 1+\lambda -\lambda h_{(i)}\right] \right\} ^{j-i-1}  \notag \\
&\times \left\{ 1-h_{(j)}\left[ 1+\lambda -\lambda h_{(j)}\right] \right\}
^{n-j}  \notag \\
&\alpha \beta \gamma (\alpha x_{(1)})^{-\beta -1}e^{-\gamma (\alpha
x_{(1)})^{-\beta }}\left[ 1+\lambda -2\lambda e^{-\gamma (\alpha
x_{(1)})^{-\beta }}\right]  \notag \\
&\times \alpha \beta \gamma (\alpha x_{(i)})h_{(i)}\left[ 1+\lambda
-2\lambda h_{(i)}\right]  \notag \\
&\times  \alpha \beta \gamma (\alpha x_{(j)})h_{(j)}\left[ 1+\lambda
-2\lambda h_{(j)}\right]
\end{align}%
where
\begin{equation*}
h_{(i)}=e^{-\gamma (\alpha x_{(i)})^{-\beta }}
\end{equation*}%
special case if $i=1$ and $j=n$ we get the joint distribution of the minimum
and maximum of order statistics%
\begin{align}\label{eq5.7}
f_{1::n:n}(x_{i},x_{j}) &=n(n-1)\text{ }\left[ F(x_{(n)})-F(x_{(1)})\right]
^{n-2}f(x_{(1)})f(x_{(n)})  \notag \\
&=n(n-1)\left\{ h_{(n)}\left[ 1+\lambda -\lambda h_{(n)}\right] -h_{(1)}%
\left[ 1+\lambda -\lambda h_{(1)}\right] \right\} ^{n-2}  \notag \\
&\times  \alpha \beta \gamma (\alpha x_{(1)})h_{(1)}\left[ 1+\lambda
-2\lambda h_{(i1}\right]  \notag \\
&\times  \alpha \beta \gamma (\alpha x_{(n)})h_{(n)}\left[ 1+\lambda
-2\lambda h_{(n)}\right] .
\end{align}

\section{Least Squares and Weighted Least Squares Estimators}

In this section we provide the regression based method estimators of the
unknown parameters of the transmuted generalized inverse Weibull
distribution, which was originally suggested by Swain, Venkatraman and
Wilson (1988) to estimate the parameters of beta distributions. It can be
used some other cases also. Suppose $Y_{1},...,Y_{n}$ is a random sample of
size $n$ from a distribution function $G(.)$ and suppose $Y_{(i)}; i=1,2,...,n$ denotes the ordered sample. The proposed method uses the
distribution of $G(Y_{(i)})$. For a sample of size $n$, we have%
\begin{align*}
E\left( G(Y_{(j)})\right) &=\frac{j}{n+1},V\left( G(Y_{(j)})\right) =\frac{%
j(n-j+1)}{(n+1)^{2}(n+2)} \\
&\text{and } Cov\left( G(Y_{(j)}),G(Y_{(k)})\right) =\frac{j(n-k+1)}{
(n+1)^{2}(n+2)};\text{for\ }j<k\text{,}
\end{align*}%
see Johnson, Kotz and Balakrishnan (1995). Using the expectations and the
variances, two variants of the least squares methods can be used.\medskip
\newline

\textbf{Method 1 (Least Squares Estimators)} . Obtain the estimators by
minimizing%
\begin{equation}\label{eq5.1}
\sum\nolimits_{j=1}^{n}\left( G(Y_{(j)}-\frac{j}{n+1}\right) ^{2},
\end{equation}%
with respect to the unknown parameters. Therefore in case of $TGIW$
distribution the least squares estimators of $\alpha ,\beta $ and $\lambda $
, say $\widehat{\alpha }_{LSE},\widehat{\beta }_{LSE}$\ and $\widehat{%
\lambda }_{LSE}$ respectively, can be obtained by minimizing%
\begin{equation*}
\sum\limits_{j=1}^{n}\left[ e^{-\gamma (\alpha x_{(i)})^{-\beta }}\left[
1+\lambda -\lambda e^{-\gamma (\alpha x_{(i)})^{-\beta }}\right] -\frac{j}{%
n+1}\right] ^{2}
\end{equation*}%
with respect to $\alpha ,$ $\beta $ and $\lambda $ .\medskip \newline

\textbf{Method 2 (Weighted Least Squares Estimators).} The weighted least
squares estimators can be obtained by minimizing%
\begin{equation}\label{eq5.2}
\sum\limits_{j=1}^{n}w_{j}\left( G(Y_{(j)}-\frac{j}{n+1}\right) ^{2},
\end{equation}%
with respect to the unknown parameters, where%
\begin{equation*}
w_{j}=\frac{1}{V\left( G(Y_{(j)})\right) }=\frac{(n+1)^{2}(n+2)}{j(n-j+1)}.
\end{equation*}%
Therefore, in case of $TGIW$ distribution the weighted least squares
estimators of $\alpha ,\beta $ and $\lambda $ , say $\widehat{\alpha }%
_{WLSE},\widehat{\beta }_{WLSE}$\ and $\widehat{\lambda }_{WLSE}$\
respectively, can be obtained by minimizing%
\begin{equation*}
\sum\limits_{j=1}^{n}w_{j}\left[ e^{-\gamma (\alpha x_{(i)})^{-\beta }}%
\left[ 1+\lambda -\lambda e^{-\gamma (\alpha x_{(i)})^{-\beta }}\right] -%
\frac{j}{n+1}\right] ^{2}
\end{equation*}%
with respect to the unknown parameters only.

\section{Maximum Likelihood Estimators}

In this section we discuss the maximum likelihood estimators (MLE's) and
inference for the $T_{GIW}$ $(\alpha ,\beta ,\gamma ,\lambda ,x)$
distribution. Let $x_{1},...,x_{n}$ be a random sample of size $n$ from $%
T_{GIW}$ $(\alpha ,\beta ,\gamma ,\lambda ,x)$ then the likelihood function
can be written as%
\begin{equation}\label{eq6.1}
L(\alpha ,\beta ,\gamma ,\lambda ,x_{_{(i)}})=\prod\limits_{i=1}^{n}\alpha
\beta \gamma (\alpha x_{(i)})^{-\beta -1}e^{-\gamma (\alpha x_{(i)})^{-\beta
}}\left[ 1+\lambda -2\lambda e^{-\gamma (\alpha x_{(i)})^{-\beta }}\right] dx
\end{equation}%
By accumulation taking logarithm of equation (6.1) , and the log- likelihood
function can be written as%
\begin{align}\label{eq6.2}
\log L &=n\ln \alpha +n\ln \beta +n\ln \gamma -(\beta
+1)\sum\nolimits_{i=1}^{n}\ln (\alpha x_{(i)})  \notag \\
&-\gamma \sum\nolimits_{i=1}^{n}(\alpha x_{(i)})^{-\beta
}+\sum\nolimits_{i=1}^{n}\ln \left[ 1+\lambda -2\lambda e^{-\gamma (\alpha
x_{(i)})^{-\beta }}\right]
\end{align}%
Differentiating equation \eqref{eq6.2} with respect to $\alpha ,\beta ,\gamma ,$ and $%
\lambda $ then equating it to zero. The normal equations become%
\begin{align}\label{eq6.3}
\frac{\partial \log L}{\partial \alpha } &=\frac{n}{\alpha }-(\beta
+1)\sum\limits_{i=1}^{n}(\frac{1}{\alpha })+\gamma \beta
\sum\limits_{i=1}^{n}x_{(i)}(\alpha x_{(i)})^{-\beta -1}  \notag \\
&-\sum\limits_{i=1}^{n}\frac{2\lambda \alpha x_{(i)}(\alpha
x_{(i)})^{-\beta -1}e^{-\gamma (\alpha x_{(i)})^{-\beta }}}{\left[ 1+\lambda
-2\lambda e^{-\gamma (\alpha x_{(i)})^{-\beta }}\right] }=0,
\end{align}%
\begin{align}\label{eq6.4}
\frac{\partial \log L}{\partial \beta } &=\frac{n}{\beta }%
-\sum\limits_{i=1}^{n}\ln (\alpha x_{(i)})+\gamma
\sum\limits_{i=1}^{n}(\alpha x_{(i)})^{-\beta }\ln (\alpha x_{(i)})  \notag
\\
&+\sum\limits_{i=1}^{n}\frac{-2\lambda \gamma e^{-\gamma (\alpha
x_{(i)})^{-\beta }}(\alpha x_{(i)})^{-\beta }\ln (\alpha x_{(i)})}{\left[
1+\lambda -2\lambda e^{-\gamma (\alpha x_{(i)})^{-\beta }}\right] }=0,
\end{align}%
\begin{equation}\label{eq6.5}
\frac{\partial \log L}{\partial \gamma }=\frac{n}{\gamma }%
-\sum\limits_{i=1}^{n}(\alpha x_{(i)})^{-\beta }+\sum\nolimits_{i=1}^{n}%
\frac{2\lambda e^{-\gamma (\alpha x_{(i)})^{-\beta }}(\alpha
x_{(i)})^{-\beta }}{\left[ 1+\lambda -2\lambda e^{-\gamma (\alpha
x_{(i)})^{-\beta }}\right] }=0,
\end{equation}%
and%
\begin{equation}\label{eq6.6}
\frac{\partial \log L}{\partial \lambda }=\sum\limits_{i=1}^{n}\frac{%
1-2e^{-\gamma (\alpha x_{(i)})^{-\beta }}}{\left[ 1+\lambda -2\lambda
e^{-\gamma (\alpha x_{(i)})^{-\beta }}\right] }.
\end{equation}%
We can find the estimates of the unknown parameters by maximum likelihood
method by setting these above nonlinear system of equations \eqref{eq6.3} - \eqref{eq6.6} to
zero and solve them simultaneously. These solutions will yield the ML
estimators for $\widehat{\alpha }$,$\widehat{\theta }$ , and $\widehat{%
\lambda }.$ For the three parameters transmuted generalized inverted
Weibull  distribution $T_{GIW}$ $(\alpha ,\beta ,\gamma \lambda ,x)$ pdf,
all the second order derivatives exist. Thus we have the inverse dispersion
matrix is given by

\begin{equation}\label{eq6.7}
\left(
\begin{array}{c}
\widehat{\alpha } \\
\widehat{\beta } \\
\widehat{\gamma } \\
\widehat{\lambda }%
\end{array}%
\right) \sim N\left[ \left(
\begin{array}{c}
\alpha \\
\beta \\
\gamma \\
\lambda%
\end{array}%
\right) ,\left(
\begin{array}{cccc}
\widehat{V_{\alpha \alpha }} & \widehat{V_{\alpha \beta }} & \widehat{%
V_{\alpha \gamma }} & \widehat{V_{\alpha \lambda }} \\
\widehat{V_{\beta \alpha }} & \widehat{V_{\beta \beta }} & \widehat{V_{\beta
\gamma }} & \widehat{V_{\beta \lambda }} \\
\widehat{V_{\gamma \alpha }} & \widehat{V_{\gamma \beta }} & \widehat{%
V_{\gamma \gamma }} & \widehat{V_{\gamma \lambda }} \\
\widehat{V_{\lambda \alpha }} & \widehat{V_{\lambda \beta }} & \widehat{%
V_{\lambda \gamma }} & \widehat{V_{\lambda \lambda }}%
\end{array}%
\right) \right] .
\end{equation}%
\begin{equation*}
V^{-1}=-E\left[
\begin{tabular}{cccc}
$V_{\alpha \alpha }$ & $V_{\alpha \beta }$ & $V_{\alpha \gamma }$ & $%
V_{\alpha \lambda }$ \\
& $V_{\beta \beta }$ & $V_{\beta \gamma }$ & $V_{\beta \lambda }$ \\
&  & $V_{\gamma \gamma }$ & $V_{\gamma \lambda }$ \\
&  &  & $V_{\lambda \lambda }$ \\
\end{tabular}%
\right]
\end{equation*}%
where
\begin{align*}
V_{\alpha \alpha } &=\frac{\partial ^{2}L}{\partial \alpha ^{2}},V_{\beta
\beta }=\frac{\partial ^{2}L}{\partial \beta ^{2}},V_{\lambda \lambda }=%
\frac{\partial ^{2}L}{\partial \lambda ^{2}},V_{\alpha \beta }=\frac{%
\partial ^{2}L}{\partial \alpha \partial \beta } \\
V_{\lambda \alpha } &=\frac{\partial ^{2}L}{\partial \alpha \partial
\lambda },
\end{align*}%
By solving this inverse dispersion matrix these solutions will yield
asymptotic variance and covariances of these ML estimators for $\widehat{%
\alpha }$,$\widehat{\beta }$ , $\widehat{\gamma }$ and $\widehat{\lambda }.$
Using \eqref{eq6.7}, we approximate $100(1-\delta )\%$ confidence intervals for $%
\alpha ,\beta ,\gamma $ and $\lambda $ are determined respectively as
\begin{equation*}
\widehat{\alpha }\pm z_{\frac{\delta }{2}}\sqrt{\widehat{V_{\alpha \alpha }}}%
,\text{ }\widehat{\beta }\pm z_{\frac{\delta }{2}}\sqrt{\widehat{V_{\beta
\beta }}}\text{\ , }\widehat{\gamma }\pm z_{\frac{\delta }{2}}\sqrt{\widehat{%
V_{\gamma \gamma }}}\text{and }\widehat{\lambda }\pm z_{\frac{\delta }{2}}%
\sqrt{\widehat{V_{\lambda \lambda }}}
\end{equation*}%
where $z_{\delta }$ is the upper $100\delta _{the}$ percentile of the
standard normal distribution.

We can compute the maximized unrestricted and restricted log-likeli\-hood functions to construct the likelihood ratio (LR) test statistic for testing on some transmuted GIW sub-models. For example, we can use the LR test statistic to check whether the TGIW distribution for a given data set is statistically \emph{superior} to the GIW distribution. In any case, hypothesis tests of the type $H_0: \theta = \theta_0$ versus $H_0: \theta \neq \theta_0$ can be performed using a LR test. In this case, the LR test statistic for testing $H_0$ versus $H_1$ is $\omega = 2(\ell(\hat\theta; x) - \ell(\hat\theta_0; x))$, where $\hat\theta$ and $\hat\theta_0$ are the MLEs under $H_1$ and $H_0$, respectively. The statistic $\omega$ is asymptotically (as $n \to \infty$) distributed as $\chi^2_k$, where $k$ is the length of the parameter vector $\theta$ of interest. The LR test rejects $H_0$ if $\omega > \chi^2_{k; \gamma}$, where $\chi^2_{k; \gamma}$ denotes the upper $100\gamma\%$ quantile of the $\chi^2_k$ distribution.

\section{Aplication}
In this section, we use a real data set to show that the TGIW distribution can be a better model than one based on
the GIW distribution. The data set given in Table 1 taken from \cite{libri} page 180 represents 50 items put into use at $t =0 $ and failure times are in weeks.

\begin{table}[hbt]
 \caption{\label{Tab1} 50 items put into use at $t =0 $ and failure times are in weeks.}
 \begin{small}
 \begin{center}
 \begin{tabular}{llllllllll}
 \hline
0.013&0.065&0.111&0.111&0.163&0.309&0.426&0.535&0.684&0.747\\
0.997&1.284&1.304&1.647&1.829&2.336&2.838&3.269&3.977&3.981\\
4.520&4.789&4.849&5.202&5.291&5.349&5.911&6.018&6.427&6.456\\
6.572&7.023&7.087&7.291&7.787&8.596&9.388&10.261&10.713&11.658\\
13.006&13.388&13.842&17.152&17.283&19.418&23.471&24.777&32.795&48.105\\
 \hline
 \end{tabular}
 \end{center}
 \end{small}
\end{table}

\begin{table}[hbt]
 \caption{\label{Tab1} Estimated parameters of the    GIW   and  TGIW distribution for 50 items put into use at $t =0 $ and failure times are in weeks.}
 \begin{small}
 \begin{center}
 \begin{tabular}{llll}
 \hline
 Model      & Parameter Estimate     & $-\ell(\cdot; x)$ \\
 \hline

  GIW& $\hat{\alpha} = 0.8537419$,$\hat{\beta} =  0.4790610$         & 168.638\\
              &  $\hat{\gamma}= 1.043654 $                 &\\\hline

TGIW& $\hat{\alpha}= 2.382715$, $\hat{\beta} = 0.5297876$       &  166.387\\
 &$\hat{\gamma}= 1.1428575$, $\hat{\lambda}= -0.7472070 $&            &          \\
      \hline
 \end{tabular}
 \end{center}
 \end{small}
\end{table}

\begin{table}[hbt]
 \caption{\label{Tab2} Criteria for comparison.}
 \begin{center}
 \begin{small}
 \begin{tabular}{lllll}
 \hline
  Model           &K-S  & $-2\ell$ & AIC     & AICC    \\
  \hline
GIW            &0.1992 &   337.276 &343.276  &343.797  \\
TGIW           & 0.1917& 332.774 & 340.774  &341.662  \\
  \hline
 \end{tabular}
 \end{small}
 \end{center}
\end{table}
The LR test statistic to test the hypotheses $H_0: \lambda = 0$ versus $H_1: \lambda \not= 0$ is $\omega = 4.502 > 3.841 = \chi^2_{1; 0.05}$, so we reject the null hypothesis.
In order to compare the two distribution models, we consider criteria like K-S, $-2\ell$, AIC (Akaike information criterion)and AICC (corrected Akaike information criterion)  for the data set. The better distribution corresponds to smaller  $-2\ell$, AIC and AICC  values:`
$$
 \mbox{AIC} = 2k - 2\ell\,, \quad \textrm{and}
 \mbox{AICC} = \mbox{AIC} + \frac{2k(k+1)}{n-k-1}\,,
$$

where $k$ is the number of parameters in the statistical model, $n$  the sample size and $\ell$ is the maximized value of the log-likelihood function under the considered model. Table 2 shows the MLEs under both distributions, table 3 shows the values of  K-S, $-2\ell$, AIC and AICC values. The values in table 3 indicate that the TGIW distribution leads to a better fit than the  GIW  distribution .

\section{Conclusion}

Here we propose a new model, the so-called the transmuted generalized inverse Weibull distribution  which extends the
generalized inverse Weibull  distribution in the analysis of data with real support. An obvious reason for generalizing a standard distribution is because the generalized form provides larger flexibility in modeling real data. We derive expansions for  moments and for the moment generating function. The estimation of parameters is approached by the method of maximum likelihood, also the information matrix is derived.  An application of TGIW distribution to real data show that the new distribution can be used quite effectively to provide better fits than GIW distribution.

\end{document}